\documentstyle[12pt]{article}
\begin{document}
\large
\begin{center}
{\bf Green's Function in Weak Interactions (in Matter) and Impossibility
of Realizing the MSW Effect}\\
\vspace{1cm}
Kh.M. Beshtoev\\
\vspace{1cm}
Joint Institute for Nuclear Research, Joliot Curie 6\\
141980 Dubna, Moscow region, Russia\\
\vspace{1cm}
\end{center}
\vspace{1cm}
\par
Abstract
\par
It is shown that the equation for Green's function of fermions (neutrinos)
with weak interactions (in matter) coincides with the equation for Green's
function of fermions in vacuum.  This result is a consequence of the fact
that the right components of fermions do not participate in weak interactions.
As a result we come to a conclusion:  the mechanism of resonance enhancement
of neutrino oscillations in matter (i.e. MSW effect) cannot exist.

\section{Introduction}
In three different approaches-by using mass Lagrangian [1, 2], by using Dirac
equation [3, 2], and using the operator formalism [4]-
I discussed the problem
of the mass generation in the standard weak interactions. The result was-
the standard weak interaction cannot generate masses of fermions since
the right components of fermions do not participate in these interactions.
Then using this result in works [4] I have shown that the effect of resonance
enhancement of neutrino oscillations in matter must not exist.
\par
At present a number of works was published (see [5] and references
there) where by using the Green's function
method it was obtained that the weak interactions can generate the
resonance enhancement of neutrino oscillations in matter (it means that the weak
interaction can generate masses).
As we see below this result is a consequence of using
weak interactions interaction term
$H^{int}_\mu = V_\mu \frac{1}{2}(1- \gamma_5)$ in inncorrect manner and
in result they obtained that right components of the fermions participate in
weak interactions.
\par
Let us consider the equation for Green's function of
fermions taking into account the standard weak interactions.

\section{Equation for Green's Function in Weak Interactions}

The Green's function method is frequently used for taking into account
electromagnetic interactions and strong interactions (chromodynamics) effects
[6]. The equation for Green's function has the following form:
$$
[\gamma^\mu (i \partial_\mu - V_\mu) - M] G(x, y) = \delta^4 (x - y) ,
\eqno(1)
$$
where $V_\mu$ characterizes electromagnetic or strong interactions and
$$
i G(x, y) = <T \Psi(x) \bar \Psi(y)>_o .
$$
It is necessary to mention that the Green's function method is a very
convenient
method for studying the electromagnetic and strong interaction effects
since these interactions are left-right side symmetric interactions.
\par
At present a  number of works was published where Green's function
was used for taking into account the weak interaction. There was shown that
the weak interaction can generate masses, i.e. masses of fermions are changed
in the weak interactions and then the resonance enhancement of neutrino
oscillations appears in matter [5]. In this work we want to show that
this result is a consequence of incorrect using a specific feature of the
standard weak interactions, namely, that the right components of fermions do
not participate in these interactions (i.e. $\Psi_R = \bar \Psi_R \equiv 0$).
\par
Usually the equation for Green's function for fermion (neutrino) with weak
interactions  is taken in the following form:
$$
[\gamma^\mu (i \partial_\mu - V_\mu) - M] G(x, y) = \delta^4 (x - y) ,
\eqno(2)
$$
where $V_\mu$ is
$$
V_\mu = V_\mu \frac{1}{2}(1 - \gamma_5) = V_\mu P_L .
\eqno(3)
$$
\par
It is supposed that the term (3) in Eq.(2) reproduces a specific feature of the
weak interactions
$$
V_\mu G(x, y) \to V_\mu \frac{1}{4}(1 - \gamma_5)^2 T(\Psi(x) \bar \Psi(y))
 = V_\mu T(\Psi_L(x) \bar \Psi_R(y)) .
$$
However, this operation is not correct since it does
not reproduce the standard weak interaction.  We see that, if we
use directly the specific feature of these interactions, then the
equation for Green's function we must rewrite in the form
$$
[\gamma^\mu (i \partial_\mu - V_\mu \left[\begin{array}{cc} \Psi_R = 0
\\ \bar \Psi_R = 0 \end{array}\right]) - M] G(x, y) = \delta^4 (x - y) ,
\eqno(4)
$$
Then the interaction term in Eq.(4) is
$$
V_\mu \left[\begin{array}{cc}\Psi_R = 0 \\ \bar \Psi_R = 0 \end{array}\right]
T(\Psi_l \bar \Psi_R (\bar \Psi_R = 0) + (\Psi_R = 0) \Psi_R \bar \Psi_L)
 = V_\mu 0 \equiv 0 ,
\eqno(5)
$$
and then Eq.(4) is transformed in the following equation:
$$
[\gamma^\mu (i \partial_\mu) - M] G(x, y) = \delta^4 (x - y) ,
\eqno(6)
$$
which coincides with the equation for free Green's function (i.e. equation
without interactions). So, we see that
the equation for Green's function with weak interactions (in matter) coincides
with the equation for Green's function in vacuum.

\section{Impossibility to Realize  the Mechanism of Resonance
Enhancement of Neutrino Oscillations in Matter}

In the previous part we have obtained that the equation for Green's function
of fermions with weak
interactions has the form (6). It is a consequence of the fact that the right
components of fermions (neutrinos) do not participate in the weak
interactions. It means that the weak interaction cannot generate masses
(see also works [1-4]) and, correspondingly, the weak
interactions do not give a deposit to effective masses of fermions (neutrinos)
therefore, the mixing angle cannot be changed in weak interactions (in matter)
and it coincides with mixing angle in vacuum. Hence, the mechanism of
resonance enhancement of neutrino oscillations in matter (MSW effect) cannot
exist.
\par
Probably, the same result takes place for renormcharge $Q^2(t)$ (where $t$ is
a transfer momentum in square) of the weak
interactions [7], i.e. renormcharge $Q^2(t)$ in the weak interactions does not
change and $Q^2(t) = const$ differs from renormcharges
$e^2(t), g^2(t)$ of the
electromagnetic and strong interactions [8].

\section{Conclusion}
It was shown that the equation for Green's function of fermions (neutrinos)
with weak
interactions (i.e. in matter) coincides with the equation for Green's function
of fermions in
vacuum.  This result is a consequence of the fact that the right components of
fermions do not participate in weak interactions. In result we have come
to a conclusion:  the mechanism of resonance enhancement of neutrino oscillations
in matter (i.e. MSW effect) cannot exist.
\par
In conclusion we would like to
stress that in the experimental data from [7] there is no visible change in
the spectrum of the $B^{8}$ Sun neutrinos. The measured spectrum of
neutrinos lies lower than the computed spectrum of the $B^{8}$
neutrinos [8]. In the case of realization of the resonance
enhancement  mechanism this spectrum must be distorted. Also, the
day-night effect on neutrinos regeneration in bulk of the Earth
keeps within the mistakes [9], i.e. it is not observed.

\par
References
\par
\noindent
1. Kh.M. Beshtoev, JINR Communication P2-93-44, Dubna, 1993.
\par
\noindent
2. Kh.M. Beshtoev, Fiz. Elem. Chastits At. Yadra (Phys.
\par
Part. Nucl.), 1996, v.27, p.53.
\par
\noindent
3. Kh.M.  Beshtoev, JINR Communication E2-93-167, Dubna, 1993.
\par
\noindent
4. Kh. M. Beshtoev, HEP-PH/9912532, 1999;
\par
Turkish Journ. of Physics (to be published).
\par
\noindent
5. C.Y. Cardall, D.J. Chung, Phys. Rev.D 1999, v.60, p.073012.
\par
\noindent
6. J. Schwinger, Phys. Rev. 1949, v.76, p.790;
\par
F.J. Dyson, Phys. Rev. 1952, v.85, p.631;
\par
N.N. Bogolubov, D.V. Schirkov, Intr. to Theory of Quant.
\par
Field, M., 1994, p.372, 437;
\par
Y. Nambu and G. Jona-Lasinio, Phys. Rev. 1961, v.122, p.345;
\par
M.K. Volkov, Fiz. Elem. Chastitz At. Yadra, 1986, v.17, p.433;
\par
A.A. Abrikosov at. al., Methods of Quant. Theory Field. in
\par
Statistic Phys. M., 1962.
\par
\noindent
7. Kh.M. Beshtoev, JINR Communication E2-94-31, Dubna, 1994.
\par
\noindent
8. T. Izikson amd K. Suber, Quant. Field Theory, M., 1984;
\par
N.N. Bogolubov, D.V. Schirkov, Intr. to Theory of Quant.
\par
Field, M., 1994, p.469.
\par
\noindent
9. K.S. Harita, et al., Phys. Rev. Lett. {\bf65}, 1297, (1991);
\par
Phys. Rev., {\bf D 44}, 2341, (1991);
\par
Y. Totsuka, Rep. Prog. Physics 377, (1992).
\par
Y. Suzuki, Proceed. of the Intern. Conf. Neutrino-98, Japan, 1998.
\par
\noindent
10. J.N. Bahcall, Neutrino Astrophysics, Cambridge U.P.
\par
Cambridge, 1989.

\end{document}